\documentclass[Physsubmission, Phys]{SciPost}

\hypersetup{
    colorlinks,
    linkcolor={red!50!black},
    citecolor={blue!50!black},
    urlcolor={blue!80!black}
}

\usepackage[bitstream-charter]{mathdesign}
\DeclareSymbolFont{usualmathcal}{OMS}{cmsy}{m}{n}
\DeclareSymbolFontAlphabet{\mathcal}{usualmathcal}
\usepackage{physics}

\def\lie{{\cal G}}
\def\a{\alpha}
\def\b{\beta}
\def\l{\lambda}
\def\cD{{\cal D}}
\def\pa{\partial}
\def\t{\tau}
\def\br{\begin{eqnarray}}
\def\er{\end{eqnarray}}

\begin{document}

\begin{center}{\Large \textbf{
	Comments on the Negative grade  KdV Hierarchy\\
}}\end{center}

\begin{center}
Y. F. Adans\textsuperscript{},
J.F. Gomes\textsuperscript{$\star$} 
G.V. Lobo\textsuperscript{} 
and
A.H. Zimerman\textsuperscript{}
\end{center}

\begin{center}
{\bf }Instituto de F\'\i sica Te\'orica, IFT-Unesp
\\
Rua Dr. Bento Teobaldo Ferraz, 271, Bloco II,\\
CEP 01140-070, S\~ao Paulo - SP, Brasil.\\

* francisco.gomes@unesp.br
\end{center}

\begin{center}
\today
\end{center}

\definecolor{palegray}{gray}{0.95}
\begin{center}
\colorbox{palegray}{
  \begin{tabular}{rr}
  \begin{minipage}{0.1\textwidth}
    \includegraphics[width=20mm]{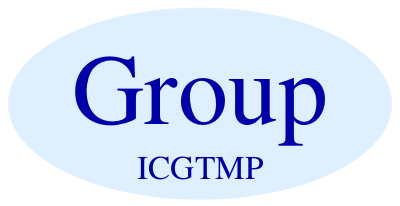}
  \end{minipage}
  &
  \begin{minipage}{0.85\textwidth}
    \begin{center}
    {\it 34th International Colloquium on Group Theoretical Methods in Physics}\\
    {\it Strasbourg, 18-22 July 2022} \\
    \doi{10.21468/SciPostPhysProc.14.014}\\
    \end{center}
  \end{minipage}
\end{tabular}
}
\end{center}

\section*{Abstract}
{\bf
The construction of negative  grade KdV hierarchy is proposed in terms of  a Miura-gauge transformation.  Such  gauge transformation  is  employed  within the zero curvature representation  and maps  the Lax operator  of the mKdV into its couterpart  within  the KdV setting.  Each odd negative KdV flow is obtained from an  odd and  its subsequent even  negative mKdV flows.  The negative KdV  flows are shown to inherit the  two 
different vacua  structure that characterizes the  associated mKdV flows.
}

\vspace{10pt}
\noindent\rule{\textwidth}{1pt}
\tableofcontents\thispagestyle{fancy}
\noindent\rule{\textwidth}{1pt}
\vspace{10pt}

\section{Introduction}
\label{sec:intro}
Integrable models have been focus of considerable  attention  in the past few years.  These are  very peculiar  two dimensional field theories admitting an infinite number of conservation laws and  soliton solutions.  
The algebraic construction  of  integrable  models has provided a series of important achievements   allowing their   construction and classification     in terms of  the decomposition of the affine algebra into graded 
subspaces.  Structural connection and  the derivation of many properties such as the construction  of  conservation laws and  soliton solutions,  can be set  from the zero curvature representation \cite{olive},\cite{babelon}.
In particular  the mKdV hierarchy, based  on the affine $sl(2)$ algebra, provides the simplest example of systematic construction  of a series of evolution equations associated to  a universal object  called Lax operator. 
 For the mKdV case  the relevant  decomposition  occurs according to 
the  principal gradation. Explicit constructions  for positive and negative graded  sub-hierarchies have been obtained. The  positive  flows are known to be labelled  by odd numbers whilst there are no restriction for the negative flows \cite{gui}.

An interesting relation between the KdV and   mKdV  hierarchies can be realised by  the Miura transformation  which maps  one hierarchy into the other. 
 In ref. \cite{ana},\cite{lobo2} we have related the two hierarchies  by a gauge transformation that maps  one   Lax operator into the other.  Such {\it Miura-gauge transformation}  acting upon the zero curvature maps  the flows from one hierarchy into the other. For the positive  sub-hierarchy the mapping is one to one, i.e.,  each flow equation  of mKdV is mapped into its counterpart  within the KdV hierarchy.    However  this is not true for the negative KdV  sub-hierarchy. In sec. 3 we argue that only  odd  flows are consistent for the KdV hierarchy and   since  there are even and odd flows within the negative mKdV  side, there  should  be  a mapping of a pair of   mKdV flows into a single KdV flow.
This is indeed true, in sect. 4 we  construct these mappings and show that  an odd and its subsequent even mKdV flows can be mapped  into a single KdV flow.  
An interesting  point to mention is that odd mKdV flows  admit only {\it zero vacuum} whilst the even admit  strictly {\it non-zero vacuum}  solutions and the associated  KdV flow  ends up inheriting  both types of structure.


\section{mKdV negative hierarchy}

In this section let us review the  construction of mKdV hierarchy  within the algebraic formalism.  Consider  the affine $ \lie =\hat {sl}(2)$ centerless Kac-Moody algebra generated by 
\begin{equation}
    h^{(m)}=\l ^{m} h^{(0)},\quad  \, E_{\pm \alpha}^{(m)}= \l ^{m} E_{\pm \a}^{(0)} \quad \text{with} \quad  \l \in \mathrm{C} \quad \text{and} \quad n \in \mathrm{Z}
\end{equation}
 satisfying  the following algebra
\begin{equation}
    \left[h^{(m)}, E_{\pm \alpha}^{(n)}\right]=\pm 2 E_{\pm \alpha}^{(m+n)}, \quad\left[E_\alpha^{(m)}, E_{-\alpha}^{(n)}\right]=h^{(m+n)}.
\end{equation}
Introduce the principal grading operator 
\begin{equation}
    {Q}_p= 2 \l \frac{d}{d\l} +\frac{1}{2} h \label{Q}
\end{equation}
that decomposes the affine algebra  into graded subspaces, i.e., $ \lie  = \bigoplus_i \lie_i$ with 
\begin{equation}
\left[Q_p, \mathcal{G}_a\right]=a \mathcal{G}_a, \qquad \left[\mathcal{G}_a, \mathcal{G}_b\right] \in \mathcal{G}_{a+b}, \quad a, b \in Z,
\end{equation}
where, for $ \lie =\hat{sl}(2)$,
\br
\mathcal{G}_{2 n} &=\left\{h^{(n)}=\lambda^n h\right\}, \qquad \qquad
\mathcal{G}_{2 n+1} &=\left\{\lambda^n\left(E_\alpha+\lambda E_{-\alpha}\right), \lambda^n\left(E_\alpha-\lambda E_{-\alpha}\right)\right\}.
\er
A second important ingredient is the choice  of  a constant   grade one element $E^{(1)} \in \lie_1$ 
\begin{equation}
    E^{(1)}=E_{\alpha}^{(0)} + E_{-\alpha}^{(1)} \label{E}
\end{equation}
such that it decomposes the affine algebra as $\hat \lie = {\cal K} \oplus {\cal {M} } $, where $\cal K$ is the \textit{Kernel} of  $E^{(1)}$:
\begin{equation} \label{Kernel}
    \mathcal{K}_E = \{  y\in {\cal {K}}, [y, E^{(1)}]=0 \} \;\; = \;\; \{ E^{(2n+1)}\equiv E_\alpha ^{(n)}+E_{-\alpha}^{(n+1)} \} \in \lie_{2n+1}
\end{equation}
 and  $\cal M$ is its complement subspace. 
 We  now  define the spatial Lax operator to be an universal  algebraic object within the  whole hierarchy to be
\begin{equation}
A_x^{\mathrm{mKdV}}(\phi ) \; =\;  E^{(1)} + A^{(0)} (\phi )
           \;  = \; E_{\alpha}^{(0)} + E_{-\alpha}^{(1)} + \pa_x\phi \, h^{(0)}
            = \mqty( \pa_x\phi & 1 \\ \lambda & - \pa_x\phi ), \label{8}
\end{equation}
where $v(x,t_{-N}) = \pa_x \phi$ is the field of the theory. We are interested in the negative time flows generated  by  the temporal Lax  operator component  of the form \cite{gui}
\begin{equation}
    A_{t_{-N}}^{\mathrm{mKdV}}= D^{(-N)}+D^{(-N+1)}+\cdots+D^{(-1)}, \qquad N=1,2, \cdots
\end{equation}
where $D^{(i)} \in \lie_i$. Thus, for a given integer $N$, the zero curvature equation 
\begin{equation}
     \left[ \pa_x + E^{(1)} + A^{(0)} , \; \pa_{t_{-N}} + D^{(-N)}+D^{(-N+1)}+\cdots+D^{(-1)} \right]  = 0
\end{equation}
decomposes according to  the grading structure, i.e.,
\begin{eqnarray}
\left[ A^{(0)}, D^{(-N)}\right]  +\pa_x D^{(-N)}  &=&0, \\
\left[A^{(0)}, D^{(-N+1)}\right]+\left[E^{(1)}, D^{(-N)}\right]+\partial_{x} D^{(-N+1)}&=& 0 ,\\
 \vdots  & & \vdots \nonumber\\
  \left[ E^{(1)}, D^{(-1)}\right]  - \pa_{t_{N}} A^{(0)}  &=&0. \label{eqmot}
\end{eqnarray}
These eqns.  can be solved  grade by grade in order to determine  $D^{(i)}$ and  the evolution equation for $A^{(0)}(\phi)$ according to  time $t_{-N}$   is given by \eqref{eqmot}. 

The  simplest case is found by taking $N=1$,  leading  to
\begin{equation}
   A_{t_{-1}}^{\mathrm{mKdV}}=    e^{-2\phi} E_{\a}^{(-1)} +e^{2\phi} E_{-\a}^{(0)} = \mqty( 0 & \lambda^{-1} e^{-2\phi} \\ e^{2\phi} & 0 ),
\end{equation}
associated with the well known sinh-Gordon equation,
\begin{equation}
    \phi_{x,t_{-1}}= e^{2\phi}-e^{-2\phi}. \label{15}
\end{equation}
 Notice that  $v = \pa_x \phi = v_0 = const. $ is the vacuum solution of (\ref{15}) only if $v_0 = 0 \rightarrow \phi =0 $.
   It therefore follows that the sinh-Gordon  equation only admits  zero vacuum solution.

Considering now  $N=2$, we find
\begin{eqnarray}
     A_{t_{-2}}^{\mathrm{mKdV}} &=& h^{(-1)}+ \left(2  e^{-2 \phi}d^{-1}(e^{2 \phi})\right) E_{\a}^{(-1)}  -2 e^{2 \phi} d^{-1}(e^{-2 \phi})  E_{-\a}^{(0)} \nonumber\\
     \nonumber\\
       &=&\left(
\begin{array}{cc}
\l ^{-1} &  \l ^{-1} \left(2  e^{-2 \phi}d^{-1}(e^{2 \phi})\right) \\
 -2 e^{2 \phi} d^{-1}(e^{-2 \phi}) & -\l ^{-1} \\
\end{array}
\right), \label{laxm2}
\end{eqnarray} 
 where we have denoted $d^{-1} f = \int_0^{x} f dx'$. It leads to the following nonlocal equation of motion
 \begin{equation} \label{eqtm2}
     \phi_{x,t_{-2}}= -2 \left(e^{-2 \phi}d^{-1}(e^{2 \phi})+e^{2 \phi}d^{-1}(e^{-2 \phi})\right). 
 \end{equation}
 Notice that   for  $\phi = \phi_0 = v_0 x$ the following identity 
 \br 
 e^{-2 v_0x}d^{-1}(e^{2 v_0x})+e^{2 v_0x}d^{-1}(e^{-2 v_0x}) = 0 
 \er 
 holds  only for $v_0 \neq 0$ and $v= v_0 $ is the vacuum solution  of (\ref{eqtm2}), only if 
 $v_0\neq 0$.
In fact, it can be shown  that  all  models associated to negative even values of $N$  only admit non-zero vacuum solutions  \cite{gui}.  
Let us consider the zero curvature equation in the vacuum regime, i.e., 
\begin{equation}
     \left[A_x^{vac}= E^{(1)} + v_0 h^{(0)} , \; A_{t_{-N}}^{vac}= D^{(-N)}_{vac}+D^{(-N+1)}_{vac}+\cdots+D^{(-1)}_{vac} \right]  = 0. \label{vacc}
\end{equation}
The lowest grade equation is 
\begin{equation}
     \left[v_0 h^{(0)} , \;  D^{(-N)}_{vac}\right]  = 0.
\end{equation}
Thus, if $v_0 \neq 0$ $D^{(-N)}_{vac}$ must commute  with $h^{(0)}$ and therefore $D^{(-N)} _{vac}\in \lie_{-2n}$  and  $N=2n$. Conversely  if $v_0 = 0$ the lowest  grade  eqn. becomes 
\begin{equation}
     \left[ E^{(1)} , \;  D^{(-N)}_{vac}\right]  = 0
\end{equation}
implying  $  D^{(-N)}_{vac} \in \mathcal{K}_E$ and $N$ is odd. Thus, the negative mKdV hierarchy splits in two sub-hierarchies: one even admitting strictly non-zero vacuum  ($v_0 \neq 0$) and one odd 
 admiting, only   zero vacuum ($v_0=0$) solutions. The systematic construction of soliton solutions for the negative mKdV hierarchies was previously studied and can be written as follows (see \cite{gui}).  
 For  the odd sub-hierarchy the one soliton solution was  constructed from    dressing the zero vacuum solution  ($ A_x^{vac} = E^{(1)}$) leading to 
\begin{equation}
    v(x,t_{-2n+1})= \pa_x \ln \left(\frac{1-\b e^{2kx+  \omega_{-2n+1}t_{-2n+1}}}{1+\b e^{2kx+  \omega_{-2n+1}t_{-2n+1}}}\right)  \quad \text{with} \quad  \omega_{-2n+1}= 2k^{-2n+1}. \label{21}
\end{equation}
For the even sub-hierarchy the constant value  of the vacuum, $v_0$ 
 introduces a deformation in the Lax operator, $ A_x^{vac} = E^{(1)}+v_0h^{(0)}$ and hence  upon the dressing method. In  \cite{gui} the solutions were constructed employing  deformed vertex operators yielding  for the one soliton solution, 
\begin{equation}
    v(x,t_{-2n})= v_0+ \pa_x \ln \left(\frac{1+\b (v_0-k) e^{2kx+ \omega_{-2n} t_{-2n}}}{1+\b (v_0+k) e^{2kx+\omega_{-2n} t_{-2n}}}\right) \quad \text{with} \quad  \omega_{-2n}= \frac{2k}{v_0(k^2 -v_0^2)^n}, \label{22}
\end{equation}
where in both cases,  $\b$ is a free parameter.


\section{KdV negative hierarchy}

For the KdV hierarchy  we employ the same algebraic structure  of section 3, i.e.,  principal gradation, ${Q}_{p}$  (\ref{Q})   and the constant grade one element $E^{(1)}$ (\ref{E}).  We propose the following Lax operator,
\begin{equation}
    A_{x}^{\text{KdV}}(J) \; = \; E^{(1)} + A^{(-1)} 
                             \; = \; E_{\a}^{(0)} + E_{-\a}^{(1)} + J \; E_{-\a}^{(0)}
                             \; = \; \mqty(0 & 1  \\
                                \lambda + J & 0) \label{23}
\end{equation}
where $A^{(-1)} =  J \; E_{-\a}^{(0)}\in \lie_{-1}$ and   $J=J(x,\tau_{N})$ is the field of KdV hierarchy. For the sub-hierarchy that leads to negative time-flow $\tau_{-N}$ ,  the  temporal-part Lax operator is given by
\begin{equation}
    A_{\t_{-N}}^{\text{KdV}}(J) = \cD^{(-N-2)} + \cD^{(-N-1)} + \cdots + \cD^{(-1)}, \label{tn}
\end{equation}
where $\cD^{(i)} \in \lie_i$. The zero curvature decomposes according to the graded structure as
\begin{eqnarray}
    \left[ A^{(-1)}, \cD^{(-N-2)} \right] &=& 0 \label{lowestkdv} \\
    \pa_x \cD^{(-N-2)} + \left[A^{(-1)}, \cD^{(-N-1)} \right] &=& 0 \\
    \pa_x \cD^{(-N-1)} + \left[E^{(1)}, \cD^{(-N-2)} \right] + \left[A^{(-1)}, \cD^{(-N)} \right] &=& 0 \\
                   &\vdots& \nonumber \\
    \pa_x \cD^{(-1)} + \left[E^{(1)}, D^{(-2)} \right] - \pa_{\t_{-N}} A^{(-1)} &=& 0 \label{eqofmotionkdv}\\
    \left[ E^{(1)}, \cD^{(-1)} \right] &=& 0, \label{nn}
\end{eqnarray}
which allows solving for all $\cD^{(i)}$ and determines the  equation of motion \eqref{eqofmotionkdv} according to $\t_{-N}$. Notice that the lowest grade equation \eqref{lowestkdv} 
  implies that   $\cD^{(-N-2)}$ is proportional to $E_{-\a}^{(-m)}$ and therefore $N = 2m-1$. For this reason {\it all equations of motion for the KdV hierarchy are associated with odd temporal flows}, 
 in contrast  to the  mKdV case, where there are  equations of motion  associated to both, even and odd flows.

The equations of motion for KdV hierarchy are  more conveniently expresed in terms of non-local field $J(x,\t_N) = \pa_x \eta(x,\t_N)$. 
The first negative flow  is obtained from zero curvature with $N=1$,  leads to the following temporal Lax operator,
\begin{eqnarray}
    A_{\t_{-1}}^{\text{KdV}} &=& \frac{\eta_{\t_{-1}}}{2} \; \left( E_{\a}^{(-1)} + E_{-\a}^{(0)}  \right)
    + \frac{\eta_{x,\t_{-1}}}{4} \; h^{(-1)}
    + \frac{ 2\eta_x\eta_{\t_{-1}} - \eta_{2x,\t_{-1}} }{ 4 } \; E_{-\a}^{(-1)} \nonumber   \\
    \nonumber \\
    &=& \mqty( \frac{ \eta_{x,\t_{-1}} }{ 4\l } & \frac{ \eta_{\t_{-1}} }{ 2\l } \\
    \frac{ 2\eta_x\eta_{\t_{-1}} - \eta_{2x,\t_{-1}} }{ 4 \l }+\frac{ \eta_{\t_{-1} }}{ 2 } & -\frac{ \eta_{x,\t_{-1}} }{ 4\l } ) \label{30}
\end{eqnarray}
and  equation of motion
\begin{equation}
    4 \eta_x \eta_{x,\t_{-1}}+2 \eta_{2x} \eta_{\t_{-1}} -\eta_{3x, \t_{-1}} = 0.
    \label{kdv:t-1}
\end{equation}
This equation was first proposed in     \cite{vero} using  the inverse of recursion operator. Later in \cite{qiao}, its Hamiltonian  and soliton solutions were discussed.

If we now take $N=3$ in (\ref{tn}) and find for the associated temporal Lax operator,
\begin{eqnarray}
    A_{\t_{-3}}^{\text{KdV}} &=& \frac{\eta_{\t_{-3}}}{2} \left( E_{\a}^{(-1)} + E_{-\a}^{(0)} \right) 
    + \frac{\eta_{x,\t_{-3}}}{4} \; h^{(-1)}
    - \frac{{\cal B}}{8} \; \left( E_{\a}^{(-2)} + E_{-\a}^{(-1)} \right) \nonumber \\
    \nonumber \\
    &+& \frac{ 2\eta_{\t_{-3}} \eta_{x} - \eta_{2x,\t_{-3} } }{ 8 } \; E_{-\a}^{(-1)} - \frac{{\cal B}_x}{16} \; h^{(-2)} + \frac{ {\cal B}_{2x} - \eta_{x} \; {\cal B} }{ 8 } \; E_{-\a}^{(-2)} \nonumber \\
    \nonumber \\
                             &=& \mqty(\frac{\eta_{x,\t_{-3}}}{4 \lambda } - \frac{{\cal B}_x}{16 \lambda ^2} & \frac{\eta_{\t_{-3}}}{2 \lambda } - \frac{{\cal B}}{8\lambda ^2} \\
 \frac{1}{2} \eta_{\t_{-3}} + \frac{2 \eta_{\t_{-3}} \; \eta_x-\eta_{2x,\t_{-3}} - {\cal B}}{8 \lambda } + \frac{ {\cal B}_{2x} - \eta_x \; {\cal B}}{8\l^2} & -\frac{\eta_{x,\t_{-3}}}{4 \lambda } + \frac{{\cal B}_x}{16 \lambda ^2}),
\end{eqnarray}
where
\begin{eqnarray}
    {\cal B} &=& d^{-1} (4 \eta_x \eta_{x,\t_{-3}}+2 \eta_{2x} \eta_{\t_{-3}} -\eta_{3x, \t_{-3}}).
\end{eqnarray}
The corresponding    equation of motion is given by
\begin{eqnarray}
       -\frac{1}{2}\eta_{5x,\t_{-3}} + 4 \eta_{x} \left( - 2 \eta_{x,\t_{-3}} \eta_{x} + \eta _{3x,\t_{-3}} - \eta_{2x} \eta_{\t_{-3}} \right) + 5 \eta _{2x} \eta _{2x,\t_{-3}} \nonumber \\
              +4 \eta _{x,\t_{-3}} \eta _{3x} + \eta _{4x} \eta _{\t_{-3}} + \eta_{2x} \; d^{-1} \left(4 \eta_x \eta_{x,\t_{-3}} + 2 \eta_{2x} \eta_{\t_{-3}} - \eta_{3x,\t_{-3}}\right)=0.
    \label{kdv:t-3}
\end{eqnarray}

Notice that vacuum solution $ \eta = \eta_0 = \text{constant}$, either zero or non-zero, satisfy both equations of motion  \eqref{kdv:t-1} and \eqref{kdv:t-3}. 
Such  behavior  differs  from the  mKdV hierarchy where the equations of motion associated with odd-time flows are satisfied with zero vacuum and the even-time flows with non-zero vacuum (constant). 
This coalescence in vacuum solution presented in KdV hierarchy can be explained more generally from zero curvature projected around vacuum, i.e,
\begin{equation}
    \left[\left. A_x^{\text{KdV}}\right\vert_{\text{vac}}, \; A_{\t_{-N}}^{\text{KdV}}\big\vert_{\text{vac}} \right]
    = \left[ E^{(1)} + \eta_0 \; E_{-\a}^{(0)}, \cD^{(-N-2)}_{\text{vac}} + \cD^{(-N-1)}_{\text{vac}} + \cdots + \cD^{(-1)}_{\text{vac}} \right] = 0.
\end{equation}
Its lowest grade component leads to 
\begin{equation}
    \left[ \eta_0 \; E_{-\a}^{(0)}, \cD^{(-N-2)}_{\text{vac}} \right] =
    \left[ \eta_0 \; E_{-\a}^{(0)}, \; a_{-N-2} \; E_{-\a}^{(-1/2(N+1))} \right] = 0,
\end{equation}
which is automatically satisfied  no matter  whether $\eta_0 $ is zero or non-zero if $N =2n-1$.  
It therefore follows that { the negative KdV hierarchy  are associated to  odd flows, $\tau_{-N} =\tau_{-2n-1}$ and admit both,  zero and non-zero vacuum solutions.}


\section{Miura Transformation and Soliton Solutions}

In order to map the mKdV and KdV hierarchies let us consider    the  {\it Miura-gauge transformation } generated by (see  \cite{ana}, \cite{lobo2} )
    \begin{equation}
    S_1 \;  = \; e^{\phi_xE_{-\a}^{(0)}} \;\; =\;\;  \mqty (1 & 0 \\ \phi_x & 1), \label{g}
\end{equation}
which maps  the two Lax operators, $ A_{x}^{\mathrm{mKdV}}$  into  $A_{x}^{\mathrm{KdV}}$ of eqns. (\ref{8}) and (\ref{23}) respectively, i.e., 
\begin{eqnarray}
     A_{x}^{\mathrm{KdV}}= S_1 A_{x}^{\mathrm{mKdV}} S_1^{-1} +S_1\partial_{x} S_1^{-1}  &=& E_{\a}^{(0)}+E_{-\a}^{(1)}+J E_{-\a}^{(0)}
\end{eqnarray}
where
\begin{equation}
    J(x,t)=\pa_x \eta(x,t) = (\phi_x)^2-\phi_{2x}.
\end{equation}
 
We now  analyse how $S_1$ acts  as a local 
gauge transformation upon $A_{t}^{\mathrm{mKdV}} $. Let us consider first  its action  on an even grade element $D^{(-2n)}= c_{-n} h^{(-n)}$:
\begin{eqnarray}
    {D}^{\mathrm{(-2n)}}& {\rightarrow} &e^{\phi_xE_{-\a}^{(0)}}\left(c_{-n}h^{(-n)}\right) e^{-\phi_xE_{-\a}^{(0)}} +e^{\phi_xE_{-\a}^{(0)}} \partial_{t} \left( e^{-\phi_xE_{-\a}^{(0)}}\right) \nonumber \\
    &=&\underbrace{c_{-n}h^{(-n)}}_{\lie_{-2n}}+\underbrace{2c_{-n}\phi_x E_{-\a}^{(-n)}}_{\lie_{-2n-1}}-\underbrace{\pa_t\phi_{x}E_{-\a}^{(0)}}_{\lie_{-1}}.
\end{eqnarray}

On the other hand, if we consider $D^{(-2n+1)}= a_{-n} E_{\a}^{(-n)}+b_{-n}E_{-\a}^{(-n+1)}$ under the local gauge generated by (\ref{g}) we find 
\begin{eqnarray}
    {D}^{\mathrm{(-2n+1)}}&\rightarrow &e^{\phi_xE_{-\a}^{(0)}}\left(a_n E_{\a}^{(-n)}+b_nE_{-\a}^{(-n+1)}\right) e^{-\phi_xE_{-\a}^{(0)}} +e^{\phi_xE_{-\a}^{(0)}} \partial_{t} \left( e^{-\phi_xE_{-\a}^{(0)}}\right) \nonumber \\
&=&-\underbrace{a_n(\phi_x)^2  E_{-\a}^{(-n)}}_{\lie_{-2n-1}}- \underbrace{a_n\phi_xh_1^{(-n)}}_{\lie_{-2n}}+\underbrace{a_nE_{\a}^{(-n)}+b_n E_{-\a}^{(-n+1)}}_{\lie_{-2n+1}} -\underbrace{\pa_t\phi_{x}E_{-\a}^{(0)}}_{\lie_{-1}}.
\end{eqnarray}
Thus, any even negative mKdV time flow of the form $A_{t_{-2n}}^{\mathrm{mKdV}}= D^{(-2n)}+D^{(-2n+1)}+\cdots + D^{(-1)}$is   mapped into its  KdV counterpart with the following graded structure,
\begin{eqnarray}
      A_{\t_{-2n+1}}^{\mathrm{KdV}}&=&e^{\phi_xE_{-\a}^{(0)}}\left(D^{(-2n)}+D^{(-2n+1)}+\cdots + D^{(-1)}\right) e^{-\phi_xE_{-\a}^{(0)}} -\phi_{x,t_{-2n}}E_{-\a}^{(0)} \nonumber\\
      &=& \mathcal{D}^{(-2n-1)}+\mathcal{D}^{(-2n)}+\cdots + \mathcal{D}^{(-1)}. \label{1}
\end{eqnarray}
For  odd negative mKdV time flow of the form  $A_{t_{-2n+1}}^{\mathrm{mKdV}}= D^{(-2n+1)}+D^{(-2n+1)}+\cdots + D^{(-1)}$  will be mapped into
\begin{eqnarray}
      A_{\t_{-2n+1}}^{\mathrm{KdV}}&=&e^{\phi_xE_{-\a}^{(0)}}\left(D^{(-2n+1)}+D^{(-2n+2)}+\cdots + D^{(-1)}\right) e^{-\phi_xE_{-\a}^{(0)}} -\phi_{x,t_{-2n+1}}E_{-\a}^{(0)} \nonumber\\
      &=& \mathcal{D}^{(-2n-1)}+\mathcal{D}^{(-2n)}+\mathcal{D}^{(-2n+1)}+\cdots + \mathcal{D}^{(-1)}. \label{2}
\end{eqnarray}
Since  $A_x^{KdV} $ is universal for both, even and odd  KdV flows,  the zero curvature representation (\ref{lowestkdv} ) - (\ref{nn}) implies that 
  $A_{t_{-2n+1}}^{\mathrm{mKdV}}$ and $A_{t_{-2n}}^{\mathrm{mKdV}}$  are transformed  by the Miura-gauge transformation (\ref{g}), into a single  graded  KdV structure 
$A_{\t_{-2n+1}}^{\mathrm{KdV}}$  (\ref{1})-( \ref{2}) (associated to flow $\t_{-2n+1}$).
 We therefore conclude that both {\it negative even and negative odd mKdV flows collapse within the same KdV odd flow}, i.e.,
\begin{equation}
    t_{-2n+1}^{mKdV}, t_{-2n}^{mKdV} \quad \underset{S_1}{\Longrightarrow} \quad \t_{-2n+1}^{KdV}.
\end{equation}
Notice that this explains why each KdV negative flow admits both zero and non-zero vacuum solutions. 
They inherit  the zero and the non-zero vacuum information from  mKdV  negative odd and its subsequent negative even    flows respectively.
Let us illustrate explicitly   for the first  two  negative mKdV flows, namely, $t_{-1}$ and $t_{-2}$. 

For $  t_{-1}^{mKdV}$ the field $\phi = \phi(x, t_{-1})$  satisfies the sinh-Gordon eqn (\ref{15}). We then have
\begin{eqnarray}
    A_{\t_{-1}}^{\mathrm{KdV}}&=&S_1 A_{t_{-1}}^{\mathrm{mKdV}} S_1^{-1} +S_1\partial_{t_{-1}} S_1^{-1} \nonumber \\
    &=&e^{\phi_xE_{-\a}^{(0)}}\left(e^{-2\phi}E_{\a}^{(-1)}+e^{2\phi}E_{-\a}^{(0)}\right) e^{-\phi_xE_{-\a}^{(0)}}  -\phi_{x,t_{-1}}E_{-\a}^{(0)},
\end{eqnarray}
leading to
\begin{equation} \label{tm11}
A_{\t_{-1}}^{\mathrm{KdV}}=e^{-2\phi}\left(E_{\a}^{(-1)}+E_{-\a}^{(0)}\right) +\frac{\eta_{x,t_{-1}}}{4} h^{(-1)}-(\phi_x)^2 e^{-2\phi}E_{-\a}^{(-1)},
\end{equation}
where we used the sinh-Gordon equation of motion, $\phi_{x,t_{-1}}= e^{2\phi}-e^{-2\phi}$ and the  \textit{Miura} transformation, $\eta_x=(\phi_x)^2- \phi_{2x}$ to simplify some terms. 
Note that in terms of zero curvature, we had already constructed   $A_{\t_{-1}}^{\mathrm{KdV}}$ given in (\ref{30}),
\begin{equation} \label{tm12}
   A_{\t_{-1}}^{\mathrm{KdV}}= \frac{\eta_{\t_{-1}}}{2} \; \left( E_{\a}^{(-1)} + E_{-\a}^{(0)}  \right)
    + \frac{\eta_{x,\t_{-1}}}{4} \; h^{(-1)}
    + \frac{ 2\eta_x\eta_{\t_{-1}} - \eta_{2x,\t_{-1}} }{ 4 } \; E_{-\a}^{(-1)}
\end{equation}
From the condition   for  eqns \eqref{tm11} and \eqref{tm12} to agree  we find
\begin{equation} \label{mt1}
    \eta_{\t_{-1}} = 2 \cdot e^{-2\phi(x,t_{-1})}. 
\end{equation}
 On the other hand, if we now consider   $t_{-2}^{mKdV}$ with $\phi = \phi(x, t_{-2})$ satisfiyng  (\ref{eqtm2}), we get from the Miura gauge transformation $A_{\t_{-1}}^{\mathrm{KdV}}=S_1 A_{t_{-2}}^{\mathrm{mKdV}} S_1^{-1} +S_1\partial_{t_{-2}} S_1^{-1} $, 
\begin{eqnarray}
   A_{\t_{-1}}^{\mathrm{KdV}}=   e^{\phi_xE_{-\a}^{(0)}}\left( h^{(-1)}+ 2  e^{-2 \phi}d^{-1}(e^{2 \phi})E_{\a}^{(-1)}  -2 e^{2 \phi} d^{-1}(e^{-2 \phi})  E_{-\a}^{(0)} \right) e^{-\phi_xE_{-\a}^{(0)}} -\phi_{x,t_{-2}}E_{-\a}^{(0)} \nonumber
\end{eqnarray}
leading to
\begin{equation} \label{tm22}
A_{\t_{-1}}^{\mathrm{KdV}}=2  e^{-2 \phi}d^{-1}(e^{2 \phi})\left(E_{\a}^{(-1)}+E_{-\a}^{(0)}\right) +\frac{\eta_{x,t_{-2}}}{4} h^{(-1)}+8(\phi_x - \phi_x^2 e^{-2\phi} d_x^{-1} e^{2 \phi} ) E_{-\a}^{(-1)}
\end{equation}
where we used the equation of motion for $t_{-2}^{\mathrm{mKdV}}$ \eqref{eqtm2} and \textit{Miura transformation}. Thus, \eqref{tm22} only agrees with \eqref{tm12} provided 
\begin{equation} \label{mt2}
    \eta_{\t_{-1}} = 2 \cdot 2  e^{-2 \phi(x,t_{-2})}d^{-1}(e^{2 \phi(x, t_{-2})}).
\end{equation}

Notice that the same $A_{\t_{-1}}^{\mathrm{KdV}}$ is written in two different  ways, one in terms of the sinh-Gordon field $\phi(x, t_{-1})$ given by (\ref{tm11})-(\ref{mt1}) 
and  another, in terms of  solution of eqn. (\ref{eqtm2}) namely $\phi(x,t_{-2})$ in (\ref{tm22})-(\ref{mt2}) .  This can be checked explicitly with solutions given in (\ref{21}) and (\ref{22}) for $n=1$.

\section{Conclusion}

We have therefore concluded from   the  above simple example that solutions  of  the KdV equation  associated  to the time flow $\tau_{-1}$  inherit  different vacuum  structures   from a pair of mKdV solutions 
 (via Miura transformation) . The first  associated  to  mKdV flow $t_{-1}$, eqn. (\ref{15})  (with zero vacuum)  satisfying  (\ref{mt1})  and the second  associated to mKdV flow $t_{-2}$, eqn. (\ref{eqtm2})  (with non-zero vacuum)  satisfying (\ref{mt2}).   The argument can be easily generalized for higher flows, and each KdV flow admits both, zero and non-zero vaccum solutions. They  are constructed from pairs  of subsequent  of  mKdV flows each of them admiting  different vacuum  structures.  We expect to report in a future publication the generalization  of our  construction to the   $A_r$ - KdV  hierarchy  employing the  gauge-Miura  transformation  proposed in \cite{lobo2}.  We  also expect to discuss the systematic construction of soliton (multisoliton) solutions and   their vacuum structure in terms of vertex operators and its deformations along  the lines of refs. \cite{gui}, \cite{ana}.
\section*{Acknowledgements}

JFG and AHZ thank CNPq and Fapesp for support. YFA thanks S\~ao Paulo Research Foundation (FAPESP) for financial support under grant $\# 2021/00623-4$ and GVL is supported by  
 Coordenação de Aperfeiçoamento de Pessoal de Nível Superior – Brasil (CAPES) – Finance Code 001.


\begin{thebibliography}{99}
\bibitem{olive} D. Olive and N. Turok, {\it Local Conserved Densities and Zero Curvature Conditions for Toda Lattice Field Theories}, { Nucl. Phys.}{\bf B215} (1983), 470, DOI: 10.1016/0550-3213(85)90347-5
\bibitem{babelon} O. Babelon, D. Bernard and  Talon, {\it{Introduction to Classical Integrable Systems}}, Cambridge Univ. Press (2009).

\bibitem{gui} J.F. Gomes, G. Starvaggi Fran\c ca, G.R. de Melo and A. H. Zimerman,   {\it Negative Even Grade mKdV Hierarchy and its Soliton Solutions}, J. Phys: Math. Theor. {\bf 42}(2009), 445204 
 arXiv:0906.5579, DOI: 10.1088/1751-8113/42/44/445204
 
 \bibitem{ana}  J.F. Gomes, A.L. Retore and A.H. Zimerman,  {\it Miura and generalized Bäcklund transformation for KdV hierarchy}, J. Phys. A: Math. Theor. {\bf49} (2016) 504003, arXiv:1610.02303,
DOI: 10.1088/1751-8113/49/50/504003
 
 \bibitem{lobo2} J. M. Carvalho Ferreira, J. F. Gomes, G. V. Lobo, A. H. Zimerman,  {\it Gauge Miura and Bäcklund transformations for generalized A n -KdV hierarchies}, J. Phys: Math. Theor. {\bf 54}(2021), 435201  arXiv:2106.00741, DOI: 10.1088/1751-8121/ac2718

 \bibitem{vero} J.M. Verosky, {\it Negative powers of Olver recursion operators}, J. Math. Phys. {\bf 32},1733 (1991), doi.org/10.1063/1.529234
 
\bibitem{qiao}Z. Qiao and Z. Fan,  {\it Negative-order Korteweg–de Vries equations}, Physical Review  E {\bf  86}, 016601 (2012), DOI: 10.1103/PhysRevE.86.016601
\end{thebibliography}

\nolinenumbers

\end{document}